\documentclass[12pt]{article}

\usepackage{amsmath,amssymb,amsfonts,amsbsy}
\usepackage{cite}
\usepackage{graphicx}
\usepackage{wrapfig}
\usepackage{epsfig}
\usepackage{bbm} % blackboard bold
\usepackage{bm} % blackboard bold
\usepackage{color}                                                       %
\usepackage{dsfont} % double stroke roman fonts                          %
\usepackage{latexsym} % a few additional special symbols                 %
\usepackage{lscape} % allows single pages to be set landscape            %
\usepackage{mathrsfs} % alternative mathematical "script" font           %
\usepackage{floatflt} % solves excess floating figure problems           %
\usepackage{slashed} % slash notation for Dirac gamma-matrix products    %
\usepackage{psfrag}

\DeclareFontFamily{OT1}{mygreek}{}%
\DeclareFontShape{OT1}{mygreek}{m}{n}{<->omsegr}{}%
\DeclareFontShape{OT1}{mygreek}{b}{n}{<->omsegrb}{}%
\DeclareFontShape{OT1}{mygreek}{m}{it}{<->omsegri}{}%
\DeclareFontShape{OT1}{mygreek}{bx}{n}{<->sub * mygreek/b/n}{}%
\DeclareFontShape{OT1}{mygreek}{m}{sl}{<->sub * mygreek/m/it}{}%
\DeclareSymbolFont{Greekrm}{OT1}{mygreek}{m}{n}
\DeclareSymbolFont{Greekbf}{OT1}{mygreek}{b}{n}
\DeclareSymbolFont{Greekit}{OT1}{mygreek}{m}{it}
\DeclareMathSymbol{\omegab}{\mathalpha}{Greekbf}{119}
\begin{document}
\addcontentsline{toc}{subsection}{{Title of the article}\\
{\it B.B. Author-Speaker}}
%%%%%%%%%%%%%
\graphicspath{{exper/yourname/}}
%or
%\graphicspath{{theory/yourname/}}
%or
%\graphicspath{{technic/yourname/}}
%%%%%%%%%%%%%

%%%%%%% please do not touch these! %%%%%%
\setcounter{section}{0}
\setcounter{subsection}{0}
\setcounter{equation}{0}
\setcounter{figure}{0}
\setcounter{footnote}{0}
\setcounter{table}{0}

\begin{center}
\textbf{THE FLAVOR STRUCTURE OF THE NUCLEON SEA}

\vspace{5mm}

\underline{J.C.~Peng}$^{\,1\,\dag}$, W.C.~Chang$^{\,2}$, 
H.Y.~Cheng$^{\,2}$ and K.F.~Liu$^{\,3}$

\vspace{5mm}

\begin{small}
  (1) \emph{Department of Physics, University of Illinois at 
  Urbana-Champaign, Urbana, Illinois 61801, USA} \\
  (2) \emph{Institute of Physics, Academia Sinica, Taipei 11529, Taiwan} \\
  (3) \emph{Department of Physics and Astronomy, University of Kentucky,
    Lexington, Kentucky 40506, USA} \\
  $\dag$ \emph{E-mail:jcpeng@illlinois.edu}
\end{small}
\end{center}

\vspace{0.0mm} % Don't laugh: it does change the spacing!

\begin{abstract}
We discuss two topics related to the flavor structure of the nucleon sea.
The first is on the identification of light-quark 
intrinsic sea from
the comparison between recent data and the intrinsic sea model
by Brodsky et al. Good agreement between the theory and data 
allows a separation of the intrinsic from the extrinsic sea
components. The magnitudes of 
the up, down, and strange intrinsic seas have been extracted. 
We then discuss the flavor 
structure and 
the Bjorken-$x$ dependence of the connected sea (CS) and disconnected
sea (DS). We show that recent data together with input from lattice
QCD allow a separation of the CS from the DS components of
the light quark sea.
\end{abstract}

\vspace{7.2mm}

\section{Introduction}

The flavor structure of the nucleon sea can provide new insight on the 
nature of QCD at the confinement scale. Perturbative QCD predicts a largely 
flavor symmetric $\bar u, \bar d, \bar s$ sea, as the 
$g \to Q \bar Q$ process, in which a gluon split into a quark antiquark 
pair ($Q \bar Q$), is insensitive to the current-quark masses of the 
$u, d,$ and $s$, which are small relative to the
QCD scale . Indeed, in the 1980s, it was commonly assumed that 
the nucleon's sea is $\bar u, \bar d, \bar s$ 
flavor symmetric,  
notwithstanding the fact that ideas based on  
meson-cloud~\cite{thomas}, Pauli-blocking~\cite{feynman}, 
and intrinsic sea~\cite{brodsky80}, already led to
predictions of a flavor asymmetric nucleon sea. We emphasize that the 
term ``flavor asymmetry" 
does not imply that any fundamental symmetry principle is violated, 
it merely refers to the differences between the $\bar u, \bar d,$ and
$\bar s$ sea quark distributions in the proton.

The earliest evidences for a flavor asymmetric nucleon sea came from 
the observation of possible violation of the Gottfried sum rule, 
suggesting $\bar u \neq \bar d$, and the charm production
in neutrino-induced deep-inelastic scattering (DIS), showing that 
strange quark sea is suppressed relative to the up and down quark 
seas. This topics continues to attract intense theoretical and 
experimental interest. We discuss some recent progress in our
understanding of the flavor structure of the nucleon sea. We first 
present the recent analysis which leads to a determination of the 
intrinsic sea components for $\bar u, \bar d$, and
$\bar s$ in the proton. We then discuss some recent effort to 
interpret the flavor structure
and momentum dependence of sea quark distributions in the context 
of connected and disconnected seas in the framework of lattice QCD. 

\section{Intrinsic versus extrinsic sea}

Brodsky, Hoyer, Peterson, and Sakai (BHPS) proposed some time ago 
that the $|u u d c \bar c\rangle$ five-quark Fock state in the proton can 
lead to enhanced production rates for charmed
hadrons at forward rapidity region~\cite{brodsky80}. 
The $c \bar c$ component in 
the $|u u d c \bar c\rangle$ configuration is coined
``intrinsic" sea in order to distinguish it from the 
conventional ``extrinsic" sea originating
from the $g \to c \bar c$ QCD process. The maximal probability 
for the $u u d c \bar c$ Fock state occurs when all five quarks move 
with similar velocities. The larger mass of the
charmed quark implies that the $c$ and $\bar c$ quarks would carry a 
large fraction of proton's
momentum. This leads to the expectation that the intrinsic charm 
has a momentum distribution which is valence-like, peaking at 
relatively large $x$. In contrast, the extrinsic
charm, which results from gluon splitting, is dominant at the 
small $x$ region. While
some tentalizing evidences for inrinsic charm have been 
reported, a study by the CTEQ 
Collaboration~\cite{pumplin} concluded that the existing 
data are not yet sufficiently 
accurate to confirm or refute the existence of intrinsic charm.

%%%%%%%% Fig. 1 %%%%%%%%
\begin{wrapfigure}[17]{R}{70mm}
%% [number of text lines to wrap]{horizontal position: LRC}{figure width}
\centering %% do not use \begin{center} ... \end{center}
\vspace*{-8mm} %% the vertical position may need tweaking
\includegraphics[width=70mm]{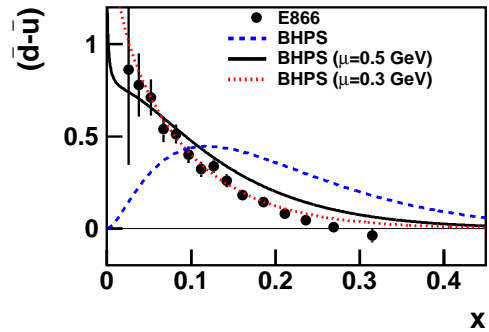}
\caption{\footnotesize
Comparison of the $\bar d(x) - \bar u(x)$ data at $Q^2 = 54$ GeV$^2$ with 
calculations. The dashed curve corresponds to
the calculation for the BHPS model, and the solid and dotted curves are
obtained by evolving it to $Q^2 = 54$ GeV$^2$ from
$\mu = 0.5$ GeV and $\mu = 0.3$ GeV, respectively.}
\label{peng_fig1}
\end{wrapfigure}
%%%%%%%%%%%%%%%%%%%%%%%%

It is natural to pose the question, ``are there any evidences 
for intrinsic sea of lighter quarks, i.e., the $|u u d u \bar u\rangle$, 
$|u u d d \bar d\rangle$, and  
$|u u d s \bar s\rangle$ Fock states?". In the BHPS 
model, the probability for the $|u u d Q \bar Q\rangle$ Fock state is expected 
to be roughly proportional to $1/m^2_Q$, where $m_Q$ is the mass 
of quark $Q$. This suggests significantly larger probabilities
for these light-quark intrinsic sea than for the intrinsic charm. 
Therefore, it is potentially easier to find evidences for these 
light-quark intrinsic sea. The challenge, however, is to come up with
ways to disentangle the intrinsic sea from the 
more abundant extrinsic sea.

In a recent attempt to search for evidences for  
intrinsic light-quark sea, two approaches were adopted 
in order to disentangle 
the intrinsic from the extrinsic sea~\cite{chang1}. The first approach 
is to select 
experimental observables which have either very little or no
contributions from the extrinsic sea. The other approach is to rely on 
the different dependences for the intrinsic and extrinsic seas. As 
mentioned earlier, the intrinsic sea is valence-like and is more 
abundant at large $x$ while the extrinsic sea is dominant at the small $x$
region.

One example of an observable free from the contribution of the
extrinsic sea is $\bar d(x) - \bar u(x)$. The perturbative
$g \to Q \bar Q$ process is expected to generate $u \bar u$
and $d \bar d$ pairs with equal probability and would have no 
contribution to $\bar d(x) - \bar u(x)$. Figure 1 shows the 
$\bar d(x) - \bar u(x)$ data from the Fermilab E866 Drell-Yan 
experiment~\cite{e866}
in comparison with the calculation using the BHPS model. The $\bar u$
and $\bar d$ are predicted to have identical $x$ dependence if
$m_u = m_d$. The exact values for the probabilities of the 
$|u u d d \bar d\rangle$ and $|u u d u \bar u\rangle$ configuration,
${\cal P}_5^{d \bar d}$ and ${\cal P}_5^{u \bar u}$, are not predicted by the 
BHPS model and must be determined from experiments. However,
the difference between ${\cal P}_5^{d \bar d}$ and ${\cal P}_5^{u \bar u}$ is
known from the moment of $\bar d(x) - \bar u(x)$, namely,

\begin{equation}
\int^{1}_{0} (\bar d(x) - \bar u(x)) dx =
{\cal P}^{d \bar d}_5 - {\cal P}^{u \bar u}_5 = 0.118 \pm 0.012,
\label{eq:eq1}
\end{equation}

\noindent where the moment is evaluated using the $\bar d(x) - \bar u(x)$
data from the Fermilab E866 experiment. Figure 1 compares the 
$\bar d(x) - \bar u(x)$ data with the calculation (dashed curve)
using the BHPS model and the constraint given by Eq. (1).
The BHPS calculation is in apparent disagreement with the 
$\bar d(x) - \bar u(x)$ data. However, the relevant scale, $\mu$,
for the BHPS model calculation is at the confinement scale, which is
much lower than the $Q^2 = 54$ GeV$^2$ scale of the E866 data. It is 
therefore necessary to evolve the BHPS result from the initial scale to 
$Q^2 = 54$ GeV$^2$. Figure 1 shows that good agreement between the
calculation (solid curve) and the data
is achieved when the initial scale is chosen as $\mu = 0.5$ GeV.
Note that an even better agreement with the data is obtained by
lowering the initial scale to $\mu = 0.3$ GeV.

%%%%%%%% Fig. 2 %%%%%%%%
\begin{wrapfigure}[21]{R}{70mm}
%% [number of text lines to wrap]{horizontal position: LRC}{figure width}
\centering %% do not use \begin{center} ... \end{center}
\vspace*{-8mm} %% the vertical position may need tweaking
\includegraphics[width=70mm]{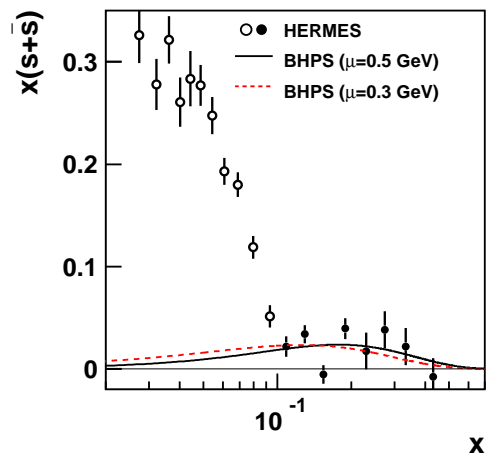}
\caption{\footnotesize
Comparison of the $x(s(x) + \bar s(x))$ data from HERMES with
calculations. The solid and
dashed curves are
obtained by evolving the BHPS result to $Q^2 = 2.5$ GeV$^2$ from $\mu
= 0.5$ GeV and $\mu = 0.3$ GeV, respectively. The normalizations of
the calculations are adjusted to fit the data at $x > 0.1$.}
\label{peng_fig2}
\end{wrapfigure}
%%%%%%%%%%%%%%%%%%%%%%%%

An example for identifying the intrinsic sea component 
by making use of its
valence-like $x$ distribution has been reported 
recently~\cite{chang2}. Figure 2
shows the extraction of $x (s(x)+ \bar s(x))$ from a measurement of
charged kaon production in semi-inclusive DIS by the HERMES 
Collaboration~\cite{hermes}.
An intriguing feature of Fig. 2 is that the strange sea falls off rapidly
with $x$ for $x<0.1$, and becomes a broad peak at the large $x$ region.
The HERMES result suggests the presence of two
distinct components of the strange sea, one at the small $x~(x<0.1)$
region and another centered at the larger $x$ region. This is in qualitative
agreement with the expectation that extrinsic and intrinsic seas have
dominant contributions at small and large $x$ region, respectively.
A comparison between the data and calculations using the BHPS model
is shown in Fig. 2 for $\mu = 0.5$ and $\mu = 0.3$ GeV. The data
at $x>0.1$ are quite well described by the calculations, 
supporting the interpretation
that the $x (s(x) + \bar s(x))$ in the valence region is dominated
by the intrinsic sea. From the normalization of the BHPS calculations
shown in Fig. 2, one can extract the probability of the $|uuds \bar s\rangle$
as 
\begin{eqnarray}
{\cal P}^{s \bar s}_5 = 0.024~~~(\mu = 0.5~\rm{GeV});~~~
{\cal P}^{s \bar s}_5 = 0.029~~~(\mu = 0.3~\rm{GeV}).
\label{eq:eq2}
\end{eqnarray}

\begin{figure}[t]
%\vspace{-1.0cm}
\begin{center}
\includegraphics[width=0.44\textwidth]{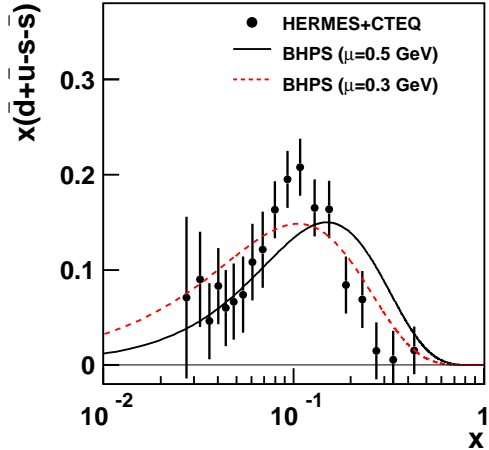}
\end{center}
\vspace{-0.3cm}
\caption{\footnotesize Comparison of $x(\bar d(x) + \bar u(x) - s(x) - \bar
s(x))$ with calculations. The solid and
dashed curves are
obtained by evolving the BHPS model calculation to 
$Q^2 = 2.5$ GeV$^2$ from $\mu
= 0.5$ GeV and $\mu = 0.3$ GeV, respectively.}
\label{peng_fig3}
\end{figure}

Another quantity which is largely free from the extrinsic sea is
$\bar u(x)+\bar d(x) -s(x)-\bar s(x)$. Under the assumption that the 
perturbative $g \to Q \bar Q$ process leads to $\bar u, \bar d, \bar s$
flavor symmetric sea, only the intrinsic sea component can contribute
to $\bar u(x)+\bar d(x) -s(x)-\bar s(x)$. From the HERMES measurement of
$x(s(x)+\bar s(x))$ and the $x(\bar u(x) + \bar d(x))$ from the 
CTEQ6.6 PDF~\cite{cteq}, we obtain $x(\bar u(x)+\bar d(x) -s(x)-\bar s(x))$
as shown in Fig. 3. We note that  
Chen, Cao, and Signal~\cite{signal} 
have also examined 
this quantity earlier in the context of the meson-cloud model. Figure 3
shows a remarkable feature that the $x(\bar u + \bar d -s - \bar s)$
distribution is valence-like and peaking at $x \sim 0.1$. One can compare
it with the BHPS model calculation 
using the following expression
\begin{eqnarray}
x(\bar u(x) + \bar d(x) - s(x) - \bar s(x)) = x(P^{u \bar u}(x_{\bar u}) +
P^{d \bar d}(x_{\bar d}) - 2 P^{s \bar s}(x_{\bar s})),
\label{eq:eq3}
\end{eqnarray}
\noindent where $P^{Q \bar Q}(x_{\bar Q})$ is the $x$ distribution 
of $\bar Q$ in the $|uud Q \bar Q\rangle$ Fock state.
Since the quantity $\bar u + \bar d - s - \bar s$ is flavor non-singlet,
it can be readily evolved from the initial scale $\mu$ to $Q^2= 2.5$ GeV$^2$.
Figure 3 shows a good agreement between the BHPS model calculation and 
the data. 
From the comparison between the BHPS calculations and the data
shown in Figs. 1-3, the probabilities for the $|uudu\bar u\rangle$,
$|uudd\bar d\rangle$, $|uuds\bar s\rangle$ Fock states can be determined
as follows (using $\mu = 0.5$ GeV):
\begin{eqnarray}
{\cal P}^{u \bar u}_5 = 0.122;~~{\cal P}^{d \bar d}_5 =
0.240;~~{\cal P}^{s \bar s}_5 = 0.024.
\label{eq:4}
\end{eqnarray}

It is remarkable that the existing data on $\bar d(x) - \bar u(x)$,
$s(x) + \bar s(x)$, and $\bar u(x) + \bar d(x) - s(x) - \bar s(x)$
not only provide a test of the BHPS model on the intrinsic sea, but also
allow an extraction of the probabilities of various five-quark
Fock states involving light antiquarks.
This result could also be extended to possible future studies on some
related topics. Some examples of these topics include:

\begin{itemize}

\item Search for intrinsic charm and beauty. From the expectation that the
probability for the $|uudQ\bar Q\rangle$ Fock state is proportional to
$1/m^2_Q$ and from the values listed in Eq. (4), one can readily estimate
that the probability
for the intrinsic charm $|uud c \bar c\rangle$ Fock state,
${\cal P}^{c \bar c}_5$, to be roughly 
$(m_s^2/m_c^2){\cal P}^{s \bar s}_5 \sim 0.003$, which is smaller than
earlier estimate~\cite{brodsky80}. Therefore, future measurements 
with higher
precision, possibly at RHIC and LHC, would be very valuable.

\item Search for intrinsic gluons. The Fock state $|uudg\rangle$ would
provide a valence-like gluon component in the proton~\cite{hoyer}. It 
remains a challenge
to identify experimental signatures for such valence-like gluons.

\item Spin and transverse-momentum dependent obsevables of intrinsic
sea. Only the spin-averaged distributions for the intrinsic sea has been
considered so far. It would be very interesting to explore
the implications of intrinsic sea on the spin-dependent and possibly
transverse-momentum dependent parton distributions of the proton.

\item Intrinsic sea for mesons and hyperons. It is straightforward to extend
the formulation for the nucleon's intrinsic sea to the cases for mesons
and hyperons. The presence of these valence-like
seas could affect, for example, the meson- or hyperon-induced Drell-Yan
cross sections in the forward rapidity region.

\item Connection between the intrinsic sea and other models. It is
important to understand the similarities and differences between the BHPS 
intrinsic sea model and other theoretical models such as the meson-cloud
model~\cite{tony} and the multi-quark model~\cite{zhang}. Some 
recent study~\cite{kfliu} has been carried out 
to elucidate the connection between the intrinsic/extrinsic seas and the 
connected/disconnected seas in the lattice QCD formulatio, as discussed
next.

\end{itemize}

\section{Connected versus disconnected sea}

In order to gain further insight on the flavor structure of the nucleon sea,
we note that, according to the path-integral formalism of the hadronic
tensor, there are two distinct sources of nucleon sea, namely, 
the connected sea (CS)
and the disconnected sea (DS). Figure 4 shows the two diagrams for the
connected  and disconnected seas. In Fig. 4 (a), the quark line propagating
backward in time between $t_1$ and $t_2$ corresponds to the connected-sea
antiquarks $\bar q^{CS} (\bar u^{CS}$ or 
$\bar d^{CS})$, which have the same flavors as the valence quarks. Figure 4 (b)
shows the DS component $q^{DS}$ and $\bar q^{DS}$ for $q = u,d,s,c$.
These two different sources of sea quarks have distinct quark flavor and 
$x$-dependence~\cite{kfliu}. While the $\bar u$ and $\bar d$ 
seas can originate from both
the CS and DS, only DS is present for the $s(\bar s)$ and $c(\bar c)$
sea. At the small-$x$ region, the CS and DS are also expected to have
different $x$ dependences. Since only reggeon exchange occurs for CS,
one expects $\bar q^{CS} \propto x^{-1/2}$ at small $x$. The presence of 
pomeron exchange implies that $\bar q^{DS} \propto x^{-1}$ at small
$x$.

\begin{figure}[tbp]
\begin{center}
\includegraphics*[width=0.25\linewidth]{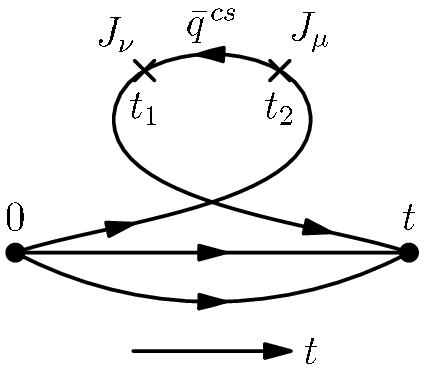}
\hspace*{8mm}
\includegraphics*[width=0.25\linewidth]{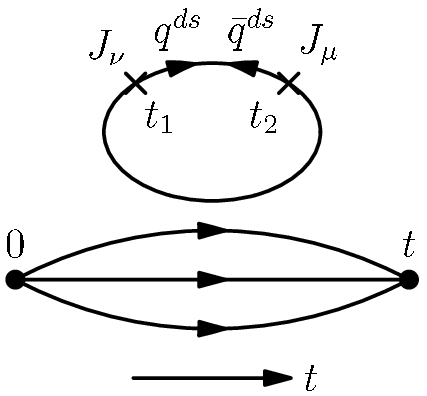}
\end{center}
\vspace*{-4mm}
\caption{\footnotesize Diagram for (a) connected sea (left) and (b)
disconnected sea (right).}
\label{peng_fig4}
\end{figure}

The distinct $x$ dependences of CS and DS remain to be checked experimentally.
Since $s$ and $\bar s$ sea is entirely originating from the DS, the
HERMES measurement of $s(x) + \bar s(x)$ provides valuable information on 
the shape of the $x$ dependence for the DS, which is not yet available
from lattice QCD calculations. The $\bar u$ and $\bar d$ seas contain
contributions from both the CS and DS. It is of interest to separate 
these two components.
A first attempt to separate the CS from the DS was reported for the
$\bar u + \bar d$ sea using the following approach~\cite{kfliu}. 
First, a plausible 
ansatz that $\bar u^{DS}(x) 
+\bar d^{DS}(x)$ is proportional to $s^{DS}(x) + \bar s^{DS}(x)$
(or equivalently, $s(x) + \bar s(x)$, since only DS contributes to
strange sea) is adopted. A recent lattice 
calculation~\cite{lattice} gives the ratio $R$ for
the moment of the strange quarks over that of the $\bar u$ plus
$\bar d$ for the disconnected diagram as
\begin{eqnarray}
R = \frac{\langle x \rangle_{s+\bar s}}{\langle x \rangle_{\bar u^{DS}
+\bar d^{DS}}} = 0.857 (40).
\label{eq:eq5}
\end{eqnarray}
\noindent Therefore, one can readily separate the CS and DS components 
for $\bar u(x) +\bar d(x)$ as follows:
\begin{eqnarray}
\bar u^{DS}(x) + \bar d^{DS}(x) = \frac{1}{R} (s(x)+\bar s(x))
\label{eq:eq6}
\end{eqnarray}
\noindent and
\begin{eqnarray}
\bar u^{CS}(x) + \bar d^{CS}(x) = \bar u(x) + \bar d(x) - 
\frac{1}{R} (s(x)+\bar s(x)).
\label{eq:eq7}
\end{eqnarray}

Figure 5 shows the decomposition of $x(\bar u(x) + \bar d(x))$
into the CS and DS components, using Eqs. (6) and (7). The $x$ dependences
for CS and DS are very different and are in qualitative
agreement with the expectation discussed earlier. This agreement lends support
to the ansatz and approach adopted in this analysis.

%%%%%%%% Fig.  %%%%%%%%
\begin{figure}[tbp]
\begin{center}
\includegraphics*[width=0.5\linewidth]{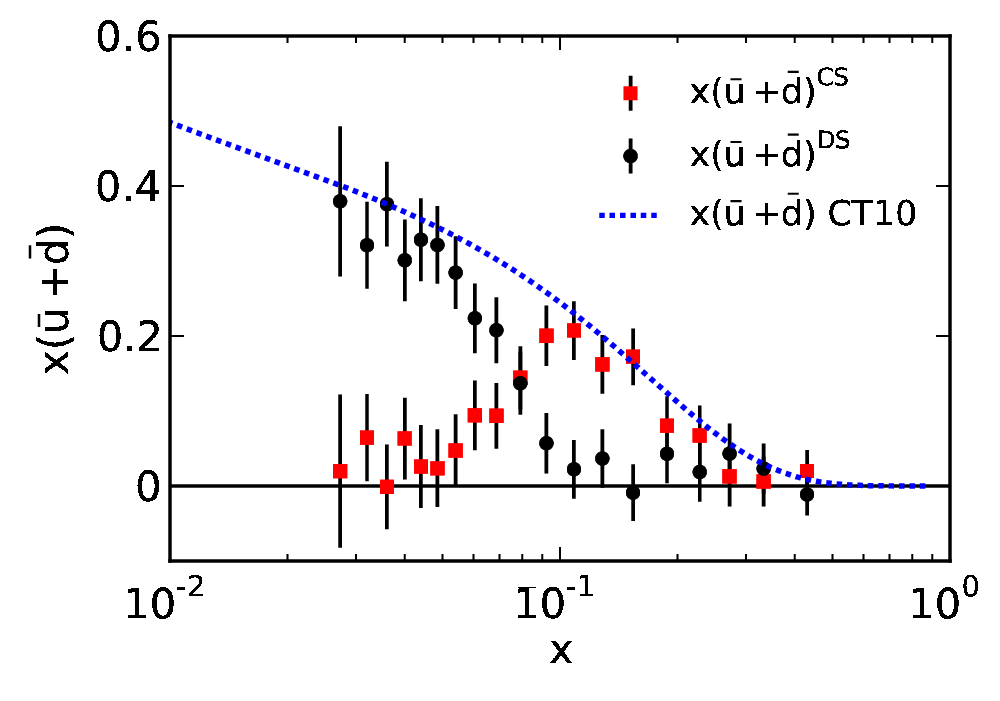}
\end{center}
\caption{\footnotesize Decomposition of $x(\bar u(x) + \bar d(x))$ into the
connected sea (CS) and the disconnected sea (DS) components using the procedure
described in the text. The CT10 parametrization of $x(\bar u(x) + \bar d(x))$
is also shown.}
\label{peng_fig5}
\end{figure}

%%%%%%%% Fig. 6 %%%%%%%%
\begin{figure}[tbp]
\begin{center}
\includegraphics*[width=0.5\linewidth]{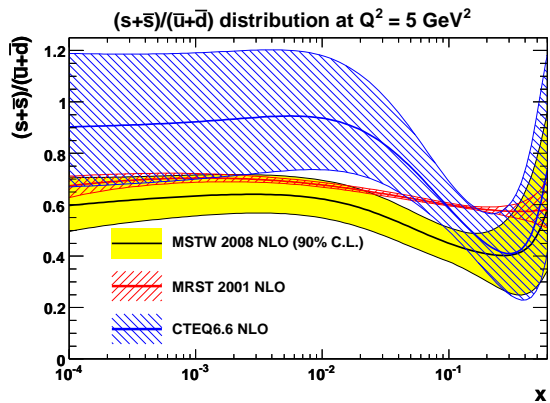}
\end{center}
\caption{\footnotesize Ratio of $s + \bar s$ over $\bar u + \bar d$ versus
$x$ at $Q^2 = 5$ GeV$^2$ from various recent
PDFs~\cite{mstw08}.}
\label{peng_fig6}
\end{figure}

From Fig. 5 one could also calculate the momentum fractions carried
by the CS and DS. It is interesting that the momentum fraction
of the $\bar u(x) + \bar d(x)$ is roughly equally divided between the 
CS and DS at $Q^2 = 2.5$ GeV$^2$. We also note that in
a very recent work~\cite{peng14}, the possible sign-change 
of $\bar d(x) - \bar u(x)$
at $x \sim 0.3$ as well as a qualitative explanation for this sign-change
in the context of lattice QCD are discussed.

The formulation of CS and DS can also explain qualitatively
the $x$ dependence of the $R(x) = (s(x)+\bar s(x))/(\bar u(x) + \bar d(x))$
ratio. Figure 6 shows the ratio $R(x)$ from some recent 
PDF sets~\cite{mstw08}.
While $R(x)$ is roughly constant at the small $x$ region, it falls with
increasing $x$ in the region $0.01 < x < 0.3$. At small $x$, the DS component
is expected to dominate, due to its $x^{-1}$ dependence. Therefore, it is
expected that $R \to 0.857$ at small $x$, according to the lattice QCD
calculation for the DS~\cite{lattice}. The recent measurement 
of $W$ and $Z$ boson productions
in $pp$ collision at 7 TeV by the ATLAS Collaboration gives
$r_s=(s+\bar s)/2\bar d$ at $x=0.023$ to 
be $1.0 + 0.25 - 0.28$~\cite{atlas}.
Both the CTEQ6.6 and ATLAS result are consistent with a roughly 
$\bar u, \bar d, \bar s$ flavor symmetric sea at small $x$. At the larger $x$
region, the valence-like CS can contribute to $\bar u$ and $\bar d$,
but not to $s$ and $\bar s$. This results in  smaller values of  
$R(x)$ at larger $x$. It is expected that future $W$ and $Z$ production data
as well as semi-inclusive kaon production data would further
improve our knowledge on the $x$ dependence of the strange quark sea.

\section{Conclusion}

In summary, we have generalized the BHPS model to the light-quark sector
and compared the model calculations with the $\bar d - \bar u$, $s + \bar s$,
and $\bar u + \bar d - s - \bar s$ data. The qualitative agreement between
the data and the calculations provides strong evidence for the existence 
of the intrinsic $u$, $d$, $s$ quark sea. This analysis also allows the
extraction of the probabilities for these Fock states. The concept of 
connected and disconnected seas in lattice QCD offers new insights on the 
flavor and $x$ dependences of the nucleon sea. Ongoing and future 
Drell-Yan (and $W/Z$ production) and semi-inclusive DIS experiments will 
provide new information on the flavor structure of the
nucleon sea. 

We acknowledge helpful discussion with Stan Brodsky and
Paul Hoyer.

\end{document}